\documentclass{IEEEtran4PSCC}
\usepackage{comment}

\usepackage{booktabs}
\usepackage{xcolor}
\definecolor{darkblue}{RGB}{0,0,135}

\newcommand{\srcom}[1]{\textcolor{black}{#1}}

\usepackage{float}
\usepackage{hyperref}
\hypersetup{
    colorlinks=true,
    linkcolor=black,
    citecolor=black,
    urlcolor=blue
}
\usepackage{cite}
\usepackage{amsmath,amssymb,amsfonts}
\usepackage{algorithmic}
\usepackage{textcomp}
\usepackage{xcolor}

\ifCLASSINFOpdf
   \usepackage[pdftex]{graphicx}
\else
   \usepackage[dvips]{graphicx}
\fi
\makeatletter
\let\old@ps@headings\ps@headings
\let\old@ps@IEEEtitlepagestyle\ps@IEEEtitlepagestyle
\def\psccfooter#1{%
    \def\ps@headings{%
        \old@ps@headings%
        \def\@oddfoot{\strut\hfill#1\hfill\strut}%
        \def\@evenfoot{\strut\hfill#1\hfill\strut}%
    }%
    \def\ps@IEEEtitlepagestyle{%
        \old@ps@IEEEtitlepagestyle%
        \def\@oddfoot{\strut\hfill#1\hfill\strut}%
        \def\@evenfoot{\strut\hfill#1\hfill\strut}%
    }%
    \ps@headings%
}
\makeatother

\begin{document}

\title{Data-Driven Koopman Predictive Control for Frequency Regulation of Power Systems using Black-Box IBRs}

\author{
\IEEEauthorblockN{Sohrab Rezaei, Xiaomo Wang, Sijia Geng}
\IEEEauthorblockA{Department of Electrical and Computer Engineering \\
Johns Hopkins University\\
Baltimore, MD, USA\\
}
}

\maketitle

\begin{abstract}
Model uncertainty of inverter-based resources (IBRs) presents significant challenges for power system control and stability. This work studies secondary frequency regulation in inverter-based power systems using a Data-driven Koopman Predictive Control (DKPC) framework. The method employs Koopman theory to lift the nonlinear system dynamics into a higher-dimensional space where they can be approximated as linear. Based on Willems' fundamental lemma, a behavioral model is constructed directly from lifted input–output data. 
A receding-horizon predictive control formulation is then provided that operates entirely using observed data, without requiring a parametric model, while satisfying explicit constraints on the control input and system output.
The proposed approach is particularly suited for IBRs with complex or uncertain dynamics. Numerical results demonstrate its effectiveness for frequency control as benchmarked against the Data-enabled Predictive Control (DeePC). The trade-off between tracking performance and control effort is illustrated through tuning of the weighting parameters.
\end{abstract}


\thanksto{\noindent Corresponding author: Sijia Geng (email: sgeng@jhu.edu).}

\section{Introduction}
Predictive control plays a central role in modern control applications due to its capability to handle multivariable systems, explicitly incorporate constraints, and optimize performance over a prediction horizon under uncertainty \cite{liu2025survey, geng2020optimal}. Its effectiveness has been widely demonstrated across diverse industrial domains. However, these methods rely on accurate system models, as both stability and performance depend directly on model fidelity. In the context of power systems, developing such accurate models is particularly challenging due to their complex, nonlinear, and uncertain dynamics. In particular, a growing number of inerter-based resources (IBRs) are being integrated to power systems, such as renewables, batteries, and FACTS devices. IBRs present further challenges in model uncertainty since there are numerous ways for control \cite{geng2025unified}, and these control designs contain proprietary information, which is rarely shared by the manufacturers with the system operators unless explicitly stipulated. This limitation motivates the exploration of predictive control strategies that can operate effectively without an accurate model and explicit system identification.

Data-driven predictive control offers a promising direction for addressing the modeling challenges of IBRs. Unlike traditional model-based methods, data-driven approaches rely solely on measured input-output data for both controller design and performance evaluation \cite{ref2}. They require no prior knowledge or assumptions about the system’s dynamics, allowing controllers to be developed directly from experimental data. While ensuring closed-loop stability and safe operation remains an active area of research, data-driven formulations provide a flexible foundation for incorporating these properties within a predictive control framework. These characteristics make data-driven predictive control suitable for IBRs.

Several works have explored data-driven predictive control for IBRs. \cite{khazaei2024data} proposes a new data-centric model identification approach for grid-connected IBRs using emerging behavioral system theories, but still relies on assumptions about the system model. Other data-driven approaches, such as model-free adaptive control \cite{shi2020data} and model-free predictive control \cite{khalilzadeh2021model}, have also been proposed for IBR modeling and control. However, these techniques face limitations, including reduced performance in certain scenarios. 

In this paper, we will develop a Data-driven Koopman Predictive Control (DKPC) for (secondary) frequency regulation of IBR-dominated power systems. 
We focus on two main ideas. The first is a behavioral model grounded in Willems' fundamental lemma (WFL), which represents the behavior of linear time-invariant (LTI) systems directly from measured input-output data without requiring an explicit parametric model \cite{ref5}. The second leverages Koopman operator theory, which maps nonlinear system dynamics into a higher-dimensional linear observable space, enabling linear analysis and control of nonlinear systems~\cite{mezic2005spectral}. 
Recent theory and algorithm developments show a converging interests: \cite{ref6} proposes a general pipeline for data-driven prediction and control in the lifted space; \cite{ref7} proves that, for nonlinear systems admitting a Koopman linear embedding, ``sufficiently wide and sufficiently deep'' data from the original system can linearly synthesize the trajectory space of the lifted system; Building on this, \cite{ref8} develop a Koopman-bilinear data-enabled predictive control (DeePC) for control-affine dynamics. These results indicate that treating nonlinear control linearly in the lifted space with data-consistent trajectory constraints is feasible and promising~\cite{rezaei2023comparative}. 
In the context of IBR applications, existing studies typically adopt DeePC or WFL without Koopman lifting (e.g., \cite{ref9} on stabilizing grid-connected converters and HVDC systems) or Koopman modeling and predictive control without WFL \cite{ref10}, \cite{ref11}. 
We explicitly couple the two threads and propose, to the best of our knowledge, the first application of behavioral predictive control in the lifted space for IBR-dominated power systems. We build an end-to-end data-driven pipeline for frequency regulation. Methodologically, we construct the lifted observations via Koopman operators, build Hankel matrices from the lifted data, and solve a receding-horizon convex program that balances tracking accuracy and control effort under hard input and output constraints. This approach avoids explicit predictor identification and mitigates multi-step rollout mismatch, and the performance is demonstrated in the IEEE 39-bus network with IBRs.

\section{Preliminaries}
To lay the groundwork for the data-driven Koopman predictive control design, this section reviews two established concepts. \textit{Willems' fundamental lemma} provides a behavioral model of LTI systems using measured input-output data with minimal assumptions, requiring only persistently exciting inputs and controllability \cite{ref5}. \textit{Koopman operator} maps nonlinear dynamics into a higher-dimensional space of observables, where their evolution is linear \cite{mezic2005spectral}. These frameworks have been combined to develop data-driven predictive control strategies capable of handling nonlinear system behavior using only measured input-output data \cite{ref6}.

\subsection{Willem's Fundamental Lemma}
Consider a discrete-time unknown LTI system $\mathcal{B}$ that generates sequences of inputs $u$ and outputs $y$. Let
\begin{equation}
\omega_k = \begin{bmatrix} u_k \\ y_k \end{bmatrix} \in \mathbb{R}^{n_u+n_y}
\end{equation}
denote the stacked input-output vector at time $k$, where $n_u$ and $n_y$ are the numbers of inputs and outputs, respectively. The Hankel matrix of depth $L$ for a sequence $\{\omega_k\}_{k=1}^{T}$ is defined as,

\begin{equation}
H_L(\omega) = \begin{bmatrix}
\omega_1 & \omega_2 & \dots & \omega_{T-L+1} \\
\omega_2 & \omega_3 & \dots & \omega_{T-L+2} \\
\vdots & \vdots & \ddots & \vdots \\
\omega_L & \omega_{L+1} & \dots & \omega_T
\end{bmatrix}.
\end{equation}

The sequence $\{\omega_k\}$ is said to be \textit{persistently exciting (PE)} of order $L$ if $H_L(\omega)$ has full row rank, assuming $T-L+1 \ge L$. 

\medskip
\noindent\textbf{Assumption 1.} System $\mathcal{B}$ is controllable in the sense of behavioral systems theory \cite{ref3}.

\noindent\textbf{Assumption 2.} The input component of $w$ is persistently exciting of order $l + n(\mathcal{B})$, where $n(\mathcal{B})$ is an upper bound on the system order and $l$ is its lag
\footnote{
In this context, the \textit{lag} of a system, denoted $l$, is the minimum number of past input-output samples needed to determine the current state.}\textcolor{blue}{\cite{ref5}}.  

\medskip

Under Assumptions 1 and 2, Willems' fundamental lemma \cite{ref5} states that for a dataset $\{u_k^d, y_k^d\}_{k=1}^{T}$, any length-$L$ sequence $\{u_k, y_k\}_{k=1}^{L}$ generated by the same system can be expressed as,
\begin{equation}
\begin{bmatrix}
H_L(u^d)\\[2pt]
H_L(y^d)
\end{bmatrix} g
=
\begin{bmatrix}
u\\[2pt]
y
\end{bmatrix},
\end{equation}

for some vector $g \in \mathbb{R}^{T-L+1}$, where:

\begin{itemize}
    \item $u^d \in \mathbb{R}^{n_u \times T}$ and $y^d \in \mathbb{R}^{n_y \times T}$ are the collected input-output data sequences of length $T$,
    \item $H_L(u^d)$ and $H_L(y^d)$ are the Hankel matrices of depth $L$ constructed from $u^d$ and $y^d$, respectively,
    \item $u \in \mathbb{R}^{n_u \times L}$ and $y \in \mathbb{R}^{n_y \times L}$ are the new input-output sequences of length $L$ to be represented as a linear combination of columns of the Hankel matrices.
\end{itemize}

\subsection{Koopman Operator}

Consider the autonomous nonlinear system,
\begin{equation}
x_{k+1} = f(x_k).
\end{equation}

\srcom{
The Koopman operator provides an alternative representation of nonlinear dynamics by shifting the focus from the state evolution to the evolution of functions of the state, referred to as observables. This perspective enables the analysis of nonlinear systems using linear tools in a lifted function space.}

Define an associated Koopman operator $\mathcal{K}$ as,
\begin{equation}
\mathcal{K}\psi = \psi \circ f,
\end{equation}
where $\psi:\mathbb{R}^{n_x}\to\mathbb{R}$ is an observable. Thus,
\begin{equation}
\mathcal{K}\psi(x_k) = \psi(f(x_k)) = \psi(x_{k+1}).
\end{equation}

\srcom{
That is, instead of directly propagating the state, the Koopman operator describes how observables evolve over time.}

A Koopman eigenfunction $\phi$ with eigenvalue $\lambda$ satisfies,
\begin{equation}
\phi(x_{k+1}) = \lambda\,\phi(x_k).
\end{equation}

Then, any observable can be decomposed as $\psi = \sum_i c_i(\psi)\,\phi_i$, where $c_i(\psi)$ is called the Koopman modes of $\psi$, and
\begin{equation}
\mathcal{K}\psi = \sum_i c_i(\psi)\,\lambda_i\,\phi_i.
\end{equation}

\srcom{
Since the exact Koopman eigenfunctions are generally not available in practice, the Koopman operator is approximated using a finite set of observables that define a lifted representation of the system. These observables map the original nonlinear dynamics into a higher-dimensional space, where the evolution can be approximated by linear dynamics \cite{williams2015data}.
In this work, the observables are constructed directly from measured outputs, enabling a data-driven lifting of the system dynamics. This lifting enables the application of Willems' fundamental lemma in the lifted space, allowing linear prediction and control of nonlinear systems without requiring an explicit model.}

\section{Problem Formulation}
In this section, we present the formulation of the data-driven Koopman predictive secondary frequency control for IBR-dominated power systems. 
The primary objective is to synthesize an optimal control policy that stabilizes the system’s frequency dynamics using only input–output data, without requiring an explicit model of the system’s nonlinear dynamics.

\subsection{Data-Driven Frequency Control of IBR-Based Systems}
Consider a power system that consists of $n$ buses and whose topology is represented by a graph $G(n, e)$, where $e\subset n \times n$ denotes the set of lines connecting the buses. Without loss of generality, assume that the first $n_\text{inv}$ buses are connected to IBRs with unknown dynamics, and the rest $n_\text{load}$ buses are connected to static loads. The system's overall dynamics are unknown due to the black-box IBR model. However, we assume that we are able to access electrical signals at the point-of-common-coupling (PCC) of each IBRs, and measure its dynamic states using, for example, a phasor measurement unit (PMU), to obtain the phase angle $\theta_i$, and estimate the angular velocity $\omega_i$. It is also practical to assume that the power outputs of the IBRs can be measured and filtered, and the IBR's power setpoint is accessible to the plant owner and dispatchable by the system operator. In the current formulation, we consider constant terminal voltage magnitudes for IBRs\footnote{In conventional power systems that are dominated by synchronous machines, when studying slower-timescale frequency response, terminal voltage can be assumed to be constant due to faster voltage control loops. In the future, we will re-evaluate such a simplification for IBR-dominated systems.}.

The (discrete-time) nonlinear dynamics of the overall system can be written in the form,
\begin{align}
x_{k+1} &= f(x_k,u_k), \\
y_k &= h(x_k),
\end{align}
where $k$ denotes the discrete time step. The state vector {$x_k \in \mathbb{R}^{2n_{\mathrm{inv}}}$} collects the dynamic states of all inverters, that is, the voltage phase agles {$\theta_i {\in \mathbb{R}}$} and frequencies {$\omega_i {\in \mathbb{R}}$}. {The input $u_k \in \mathbb{R}^{n_{u}}$ represents the vector of control inputs, where each component corresponds to an external adjustment to the active power setpoint $p_i^*$ of an individual IBR for secondary frequency control in response to power disturbances.} The output {$y_k \in \mathbb{R}^{n_{y}}$} corresponds to the inverter's frequencies. The functions $f(\cdot)$ and $h(\cdot)$ denote the (unknown) nonlinear state transition and measurement mappings, respectively.

\subsection{Data-Driven Koopman-Based Predictive Control}
\srcom{To obtain a finite-dimensional approximation of the Koopman operator, we construct the lifting using radial basis function (RBF) observables. Specifically, we consider a set of basis functions $\psi_i(\cdot)$ for $i=1,\dots,n_{\text{basis}}$, where $n_{\text{basis}}$ denotes the number of RBF observables used in the lifting. Each $\psi_i$ is a scalar observable of the output and contributes one lifted coordinate,}
\begin{equation}
\psi_i(y_k)=\|y_k-c_i\|_2^2\cdot\log_{10}\left(\|y_k-c_i\|_2\right),\label{eq:RBF}
\end{equation}
\srcom{where $c_i$ is the center associated with the $i$-th basis function, selected randomly, and $\|\cdot\|_2$ denotes the Euclidean norm.
The lifted representation is constructed by evaluating all basis functions at the current output $y_k$. In this construction, the RBF functions define a finite-dimensional observable space in which the nonlinear system behavior can be approximately represented using linear dynamics.}

\noindent\textbf{Assumption 3.} The lifted system constructed using {observable basis functions} \eqref{eq:RBF} is controllable in the behavioral systems theory\srcom{\cite{ref7,ref6}}. 

Based on Willems' fundamental lemma, if a sequence $\{z_k,u_k,y_k\}_{k=1}^{T}$ is available with lifted states $z_k=\psi(y_k)$, where $\psi:\mathbb{R}^{{n_y}}\to\mathbb{R}^{{n_{\text{basis}}}}$ is a set of {observable basis functions}, and $u$ is persistently exciting of order {{$n_{\text{basis}}+ L$}}, then any valid length-$L$ trajectory $\{z_k,u_k,y_k\}_{k=1}^{L}$ can be {represented as a linear combination},
\begin{equation}
\begin{bmatrix}
Z\\ U\\ Y
\end{bmatrix} g
=
\begin{bmatrix}
z_{[t-T_{\mathrm{ini}}+1:t+N]}\\[3pt]
u_{\mathrm{ini}}\\[1pt]
u\\[1pt]
y_{\mathrm{ini}}\\[1pt]
y
\label{eq:dd_model}
\end{bmatrix},
\end{equation}
where $Z$, $U$, and $Y$ are the depth-$L$ Hankel matrices constructed from the lifted data, system input, and output data, respectively. $T_{\mathrm{ini}}$ (chosen such that $T_{\mathrm{ini}} \ge l$) {denotes the length of the initial trajectory used to initialize the prediction}, while $N$ represents the prediction horizon and $L=T_{\mathrm{ini}}+N$. The sequences $u_{\mathrm{ini}}$ and $y_{\mathrm{ini}}$ correspond to the initial control inputs and outputs, each of length $T_{\mathrm{ini}}$, while $u$ and $y$ are the future components of length $N$ and are part of the decision variables in the optimization \eqref{optimization_problem}. \srcom{
The lifted representation enables the application of Willems' fundamental lemma in the observable space, allowing predictive control to be performed directly from data without requiring an explicit model.}

We formulate the following convex program for the synthesis of the data-driven Koopman-based predictive frequency control, considering a time horizon of $N$ steps, 

\begin{align}
\min_{u,y,z,g}\quad   \sum_{k=1}^N & \left( \|y_k-r_{t+k}\|_Q^2 + \|u_k-{u^s_{t+k}}\|_R^2 \right) + \lambda_g \|g\|_2^2 \label{optimization_problem} \\
\text{subject to}\quad 
& (\ref{eq:dd_model}), \nonumber \\
& z_k(t)=\psi(y_k(t)),\quad \forall k\in[1,N], \nonumber  \\
& u_k \in {\mathcal{U}},\quad \forall k\in[1,N], \nonumber  \\
& y_k \in {\mathcal{Y}},\quad \forall k\in[1,N].\nonumber 
\end{align}

\noindent

Here, $t$ is the current step at which the optimization problem is solved, and $r_{t+k}$ denotes the desired reference at step $t+k$. The weighted norms $\|\cdot\|_Q$ and $\|\cdot\|_R$ represent the relative importance of output tracking and control effort, respectively, with positive definite weighting matrices $Q$ and $R$. The first term in the objective penalizes deviations of the predicted outputs $y_k$ from the reference trajectory $r_{t+k}$, which, in the frequency regulation problem, typically corresponds to the nominal frequency setpoint. The second term penalizes deviations of the control input $u_k$ from the nominal input {$u^s_{t+k}$}. Finally, the regularization term $\lambda_g \|g\|_2^2$ improves numerical conditioning and ensures robustness and stability of the optimization solution by discouraging excessively large coefficients in $g$.
 
The constraints ensure that the predicted input-output sequence can be represented by the data using the learned Koopman dynamics, and that inputs and outputs stay within physical limits {$\mathcal{U}$ and $\mathcal{Y}$, respectively.

At each time step, the optimization problem is solved, and the first control input from the optimal control sequence $\bar{u}$ is applied to the nonlinear system. The new output is measured and appended to the dataset. This process is repeated at each step until the end of the simulation. 
\srcom{
Moreover, when appropriate regularization or terminal ingredients are included, practical closed-loop stability of the resulting controller can be established under bounded Koopman approximation error\footnote{This is assuming the finite-dimensional Koopman lifting constructed using the basis functions in \eqref{eq:RBF} provides a sufficiently accurate surrogate model of the system dynamics over the prediction horizon, such that the prediction error of the lifted system remains bounded and the resulting predictive control problem admits a stabilizing solution.}, as shown in~\cite{worthmann2024data,schimperna2025koopman_mpc_stability}.}

\section{System Model for Data Generation}
To validate the proposed method, we conduct a simulation study using the IEEE 39-bus network to generate the data for control design. The system contains ten machines, multiple loads, and other components, and the detailed description is available in \cite{athay1979practical}. We replace the ten machines with ten grid-forming (GFM) IBRs with the same generation level.
For our purposes, we considered a droop-based GFM model as described in \cite{chatterjee2024effects}, but ignored the inner loops, and discretized the time. 
The equations governing the frequency deviation for the $i$-th inverter, $\omega_i$, and its angle $\theta_i$ are given by,
\begin{align}
\omega_i[k+1] &= K_p(p_i^*+\tilde{u}_i[k+1]-\tilde{p}_i[k+1]),\label{eq:droop}\\
\theta_i[k+1] &= \theta_i[k]+\Delta t\mathrm{~}\omega_i[k],
\end{align}
where \eqref{eq:droop} is the frequency droop equation. $K_p$ is the active power droop coefficient.
$p_i^*$ is the active power setpoint of the inverter. $\tilde{p}_i$ is the filtered active power measurement, and $\tilde{u}_i$ is the filtered external control input. They are expressed as,
\begin{align}
\tilde{u}_i[k+1]&=\tilde{u}_i[k]+\Delta t\cdot\omega_{pc}({u}_i[k]-\tilde{u}_i[k]),\\
\tilde{p}_i[k+1] &= \tilde{p}_i[k]+\Delta t\cdot\omega_{pc}\big(p_i[k]-\tilde{p}_i[k]\big),
 \end{align}
where ${u}_i$ is the external control input. $p$ is the active power injection.
$\omega_{pc}$ is the 3dB cut-off frequency of the low-pass filter.

The active power injection $p$ relates to the neighboring buses' angles and line parameters as,
\begin{align}
p_i[k] &= \sum_{j\neq i}B_{ij}\sin\big(\theta_i[k]-\theta_j[k]\big),
 \end{align}
where $\theta_i$ and $\theta_j$ are the angles at buses $i$ and $j$, respectively, and $B_{ij}$ denotes the line admittance between buses $i$ and $j$.

We use this model solely to simulate system trajectories and generate data for control synthesis, and not in the controller design itself. 
 The fact that the predictive controller is built using only the measured input-output trajectories, following the Willemsian behavior paradigm, ensures that our control approach does not rely on knowledge of the system equations.

\section{Simulation Studies}
This section presents the numerical studies and results of the proposed method on the IBR-based IEEE 39-bus system. 

\subsection{Dataset Generation} 
To ensure the data informativity for behavioral control, that is, persistently exciting, sufficiently long samples and random inputs will be used.
A dataset is generated by applying uniformly distributed random inputs $u_{d}\in [-1,1]$ over a horizon of $T = 1000$ time steps. 
The initial active power setpoint $p^*$ and other parameters for each GFM inverters \cite{chatterjee2024effects} \cite{geng2025unified} are listed in Table~\ref{table_1}.

\begin{table}[!ht]
\renewcommand{\arraystretch}{1.3}
\centering
\caption{ Initial Value and Parameters for the GFM Inverters}
\label{table_1}
\begin{tabular}{|c|c|c|c|c|}
\hline
Parameter  & $p^*$ & $\omega_b$ & $K_p$ & $\omega_{pc}$  \\
\hline
Nominal value & 1 p.u. & 1 p.u. & 7\% & 332.8 rad/s  \\
\hline
\end{tabular}
\end{table}

For each time step, the nonlinear dynamical system simulator is used to compute the next states $\theta_{k+1}$, $\omega_{k+1}$ based on current states and inputs. The outputs $y_d$ are chosen to be the angular frequencies $\omega_i$ of the inverters, which are the variables to be regulated. 
Hankel matrices are constructed from the past and future trajectories of both the original outputs and the lifted observables. These matrices are fundamental to the data-driven predictive control framework and approximate the Koopman operator dynamics in a lifted space. They are provided to the optimization problem for control synthesis.

\subsection{Receding-Horizon Predictive Control}
The predictive controller has a prediction horizon {$N=10$}, and the simulation runs for $T_\text{sim} = 150$ time steps {with a time step size of {$\Delta t = 0.01$ s}}. At each time step, a convex problem is solved using CVX. {The lifted observable is constructed using $n_\text{basis} = 40$ RBF kernels.}
The modest lifting dimension and local RBF kernel keep the problem size tractable. 
{The weighting matrices are chosen as $Q = 3\times10^{2}I_{10}$ and $R = 10^{-2}I_{10}$, where $Q$ penalizes the frequency tracking error and $R$ penalizes the control effort. Here, $I_{10}$ denotes the 10-dimensional identity matrix, since the system contains 10 inverter control inputs and 10 corresponding frequency outputs. The regularization parameters are set as $\lambda_g = 500$, where $\lambda_g$ weights the coefficient vector $g$.}

\subsection{Numerical Results on Control Performance}

\begin{figure}[!ht]
  \centering
    \includegraphics[width=\linewidth]{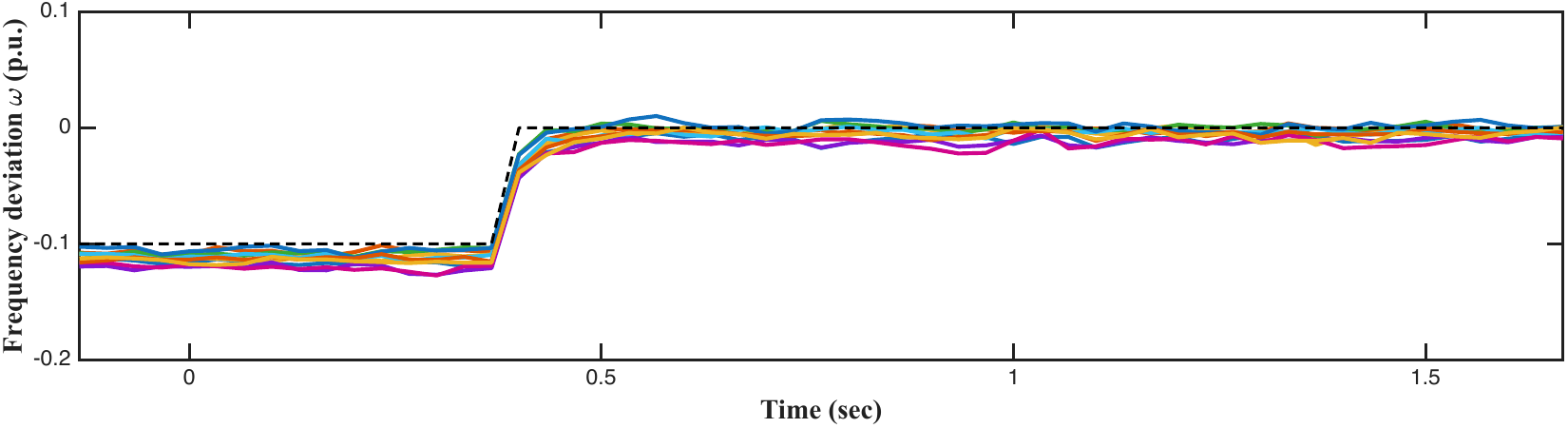}
     \includegraphics[width=\linewidth]{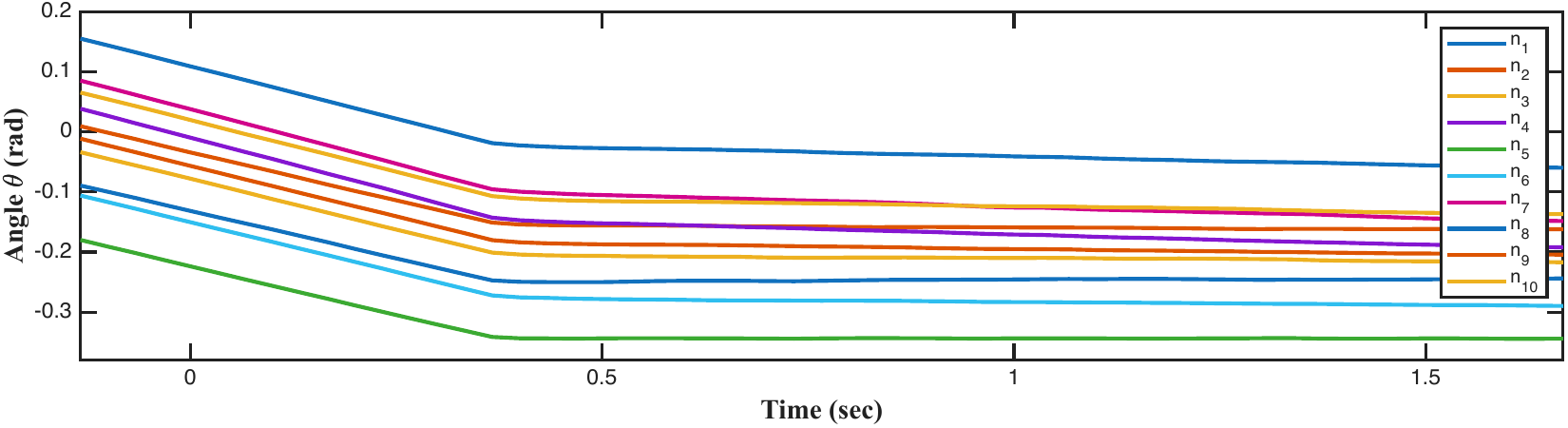}
    \caption{Frequency deviation and angle of the ten inverters in the IBR-based IEEE 39-bus system.}
  \label{fig_sim_w}
\end{figure}

Figure~\ref{fig_sim_w} shows the frequency response (upper subplot) and angle dynamics (lower subplot) for the ten inverters. Initially, the system is in a disturbed condition, where the frequency deviation is non-zero. At $0.4$ sec, the data-driven controller starts to take effect, and as can be seen, the controller was effective in bringing the frequency deviation back to the nominal value, and the tracking error in frequency deviations gradually decays.
Furthermore, the frequency tracking during the transient is smooth, with negligible overshoot, and has a short settling time. This demonstrates good performance of the controller. As can be seen in the lower subplot, the network-level synchronization is maintained. After the frequency deviation is restored to zero, the angles stop evolving, indicating that the initial disturbance is effectively suppressed and the nominal operating point is recovered. 

\begin{figure}[!ht]
  \centering
    \includegraphics[width=\linewidth]{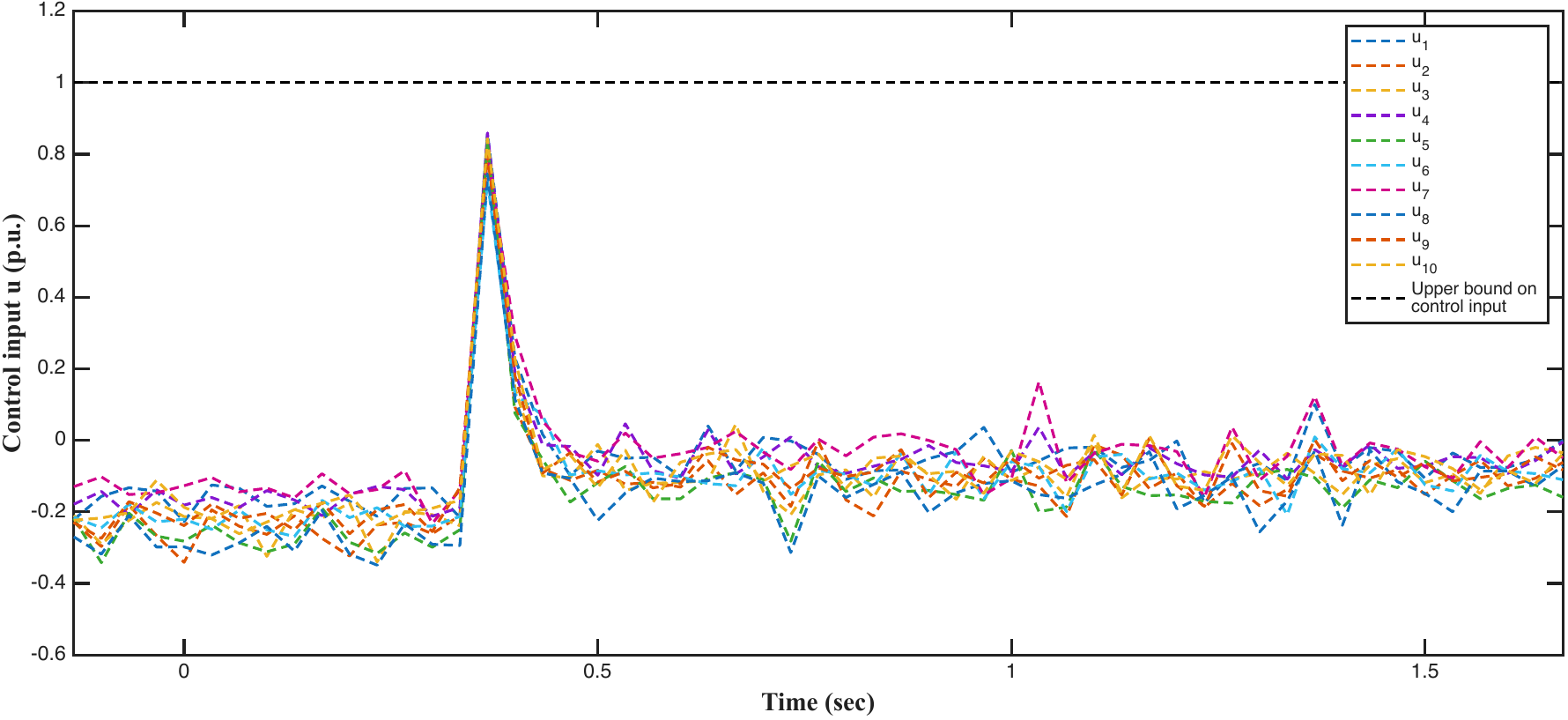}
    \caption{Control inputs of the ten inverters in the IBR-based IEEE 39-bus system.}
  \label{fig_sim_control}
\end{figure}

The control inputs for the ten inverters are plotted in Fig.~\ref{fig_sim_control}. Immediately following the reference change, the control efforts experienced a large increase to achieve fast frequency control. However, they remained strictly within the hard bounds of $[-1,1]$ on control effort, and after the initial transient period, quickly returned to relatively small values. The control signals are relatively smooth, with only minor jitters, suggesting a well-posed optimization and sufficiently rich RBF-based lifting for capturing nonlinear behavior. Moreover, the control actions of the inverters exhibit coordinated responses during the transient, while remaining within admissible limits, highlighting interoperability across the network.

\subsection{Impacts of weights on tracking and control costs}

\textcolor{black}{In this section, we study the impacts of weights on tracking error and control efforts. Here, the tracking weight matrix and control effort weight matrix are chosen as $Q = qI_{10}$ and $R = rI_{10}$, respectively, where $q$ and $r$ are scalar tuning parameters and $I_{10}$ is the identity matrix of dimension 10. Thus, the relative trade-off between frequency tracking and control effort is characterized by the ratio $q/r$. Figs.~\ref{fig_Q} and \ref{fig_U} display the frequency deviation and control inputs of the inverters under various values of $q/r$. As can be seen in Fig.~\ref{fig_Q}, increasing $q/r$, corresponding to a higher emphasis on tracking performance relative to control effort, leads to smaller transient error. The corresponding control efforts are plotted in Fig.~\ref{fig_U}, which shows stronger actuation with higher $q/r$. The quantitative performance comparison is provided in Fig.~\ref{fig_Performance}. The left subplot shows the trend of tracking performance as measured by the integrated time-weighted absolute error (ITAE)\cite{graham1953}, denoted by $\epsilon$ and defined as 
\begin{equation}
\epsilon = \sum_{k=1}^{T} t_k \, |e_k|,\label{eq:epsilon}
\end{equation}
where \(e_k\) is the frequency tracking error at time step \(k\) and \(t_k\) is the corresponding time instant. 
The right subplot shows the control effort $J_u$, defined as
\begin{equation}
J_u = \sum_{k=1}^{T} \|u_k\|_2,\label{eq:Ju}
\end{equation}
where \(u_k\) is the control input vector at time step \(k\).}

\begin{figure}[!ht]
  \centering
  \includegraphics[width=\linewidth]{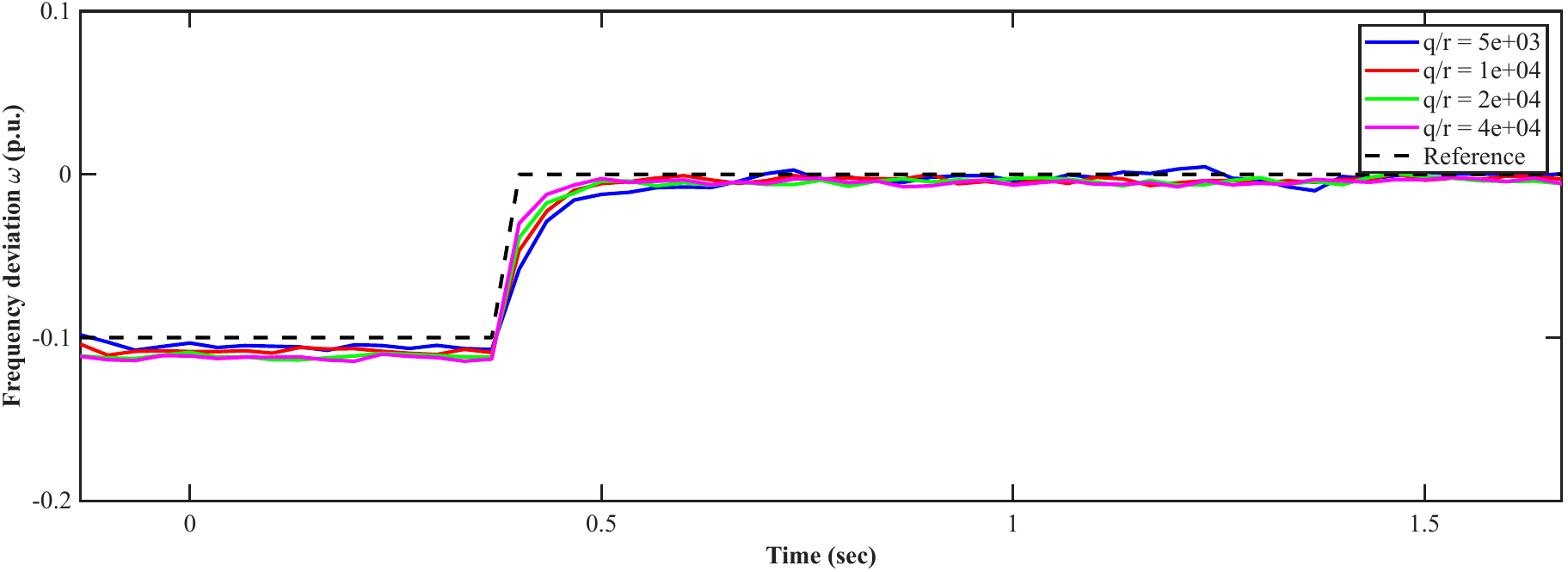}
    \caption{Frequency deviation of inverters under various $q/r$ ratios.}
  \label{fig_Q}
\end{figure}

\begin{figure}[!ht]
  \centering
    \includegraphics[width=\linewidth]{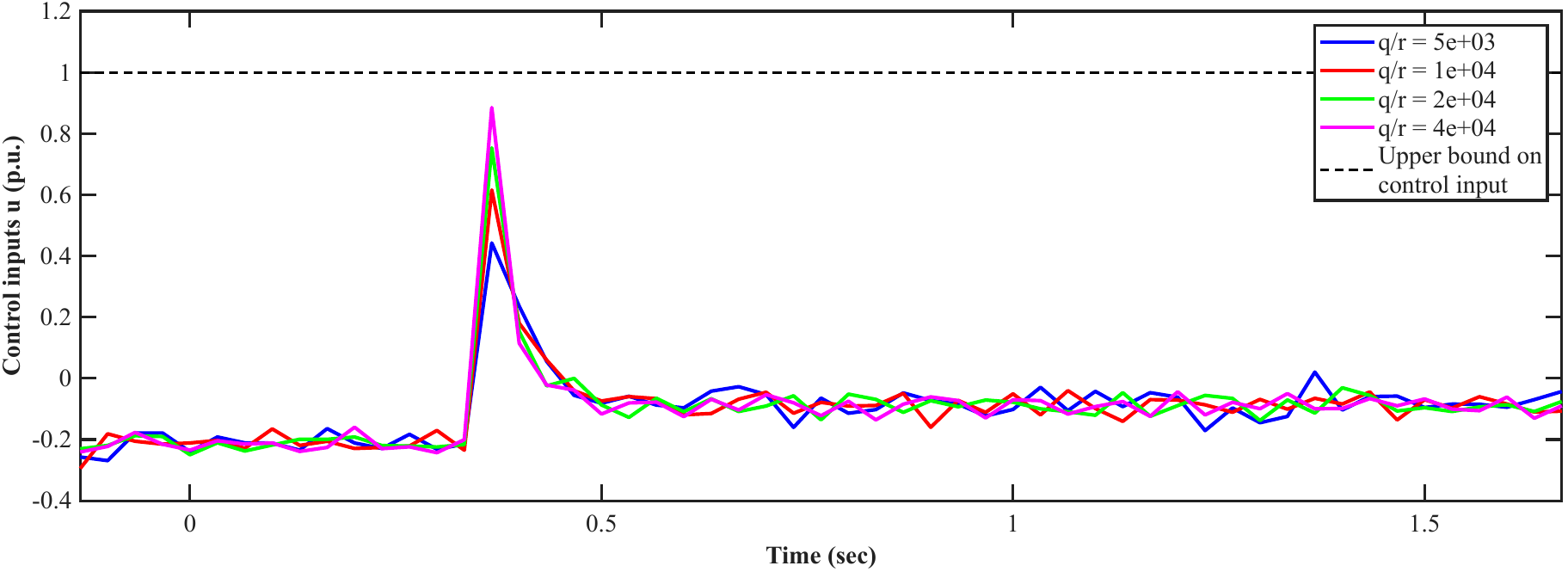}
    \caption{Control inputs of inverters under various $q/r$ ratios.}
  \label{fig_U}
\end{figure}

\begin{figure}[!ht]
  \centering
\includegraphics[width=\linewidth]{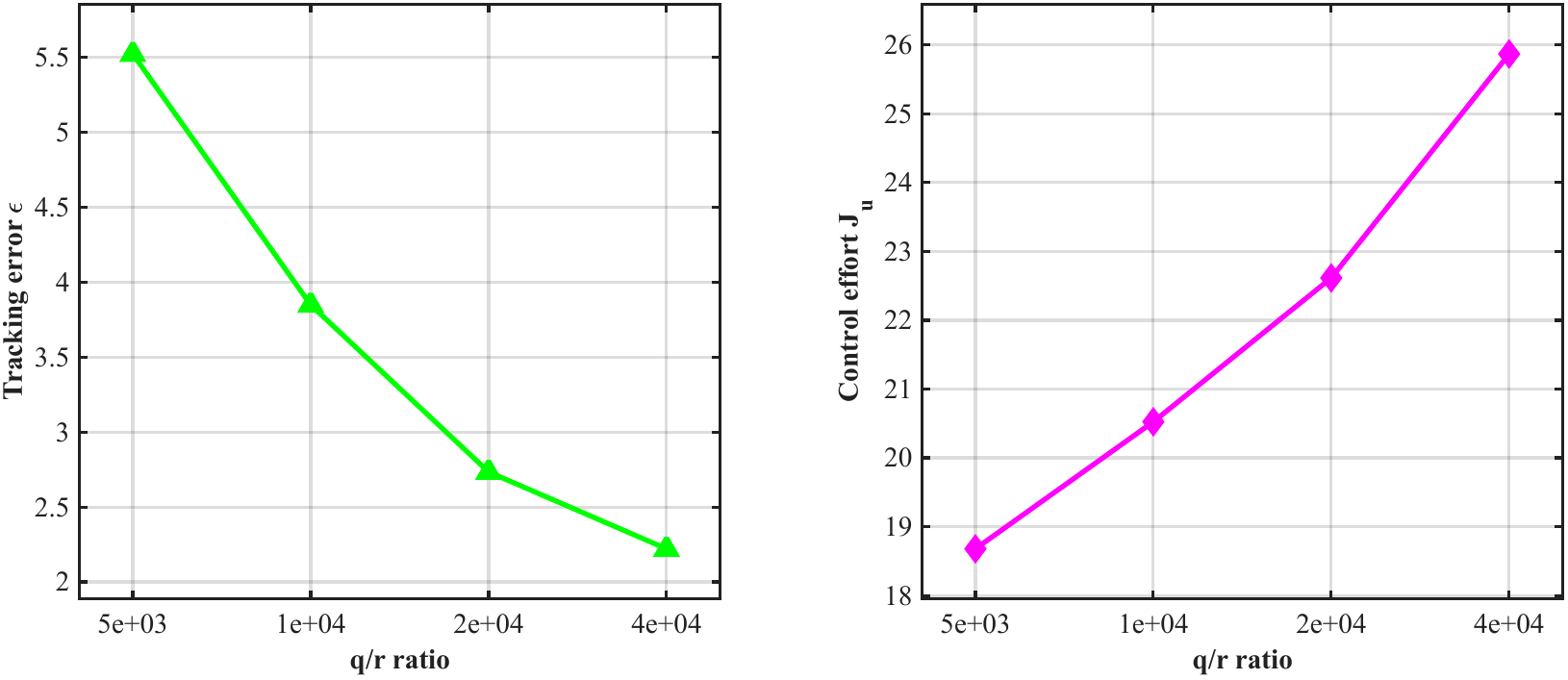}
    \caption{Quantitative performance comparison under various $q/r$ ratios.}
  \label{fig_Performance}
\end{figure}

\subsection{Comparison with Data-Enabled Predictive Control}\label{subsec:DKPC_deepc_comparison}

{\color{black}To evaluate the effectiveness of the DKPC framework, we compare its performance with Data-enabled Predictive Control (DeePC), a widely studied data-driven predictive control approaches based on Willems’ fundamental lemma. DeePC constructs predictive controllers directly from measured input–output trajectories without requiring an explicit parametric model and has been applied to power electronic systems and grid-connected converters \cite{ref9}. This provides a suitable benchmark for the proposed method, and helps determine whether the Koopman lifting provides practical benefits in terms of tracking performance and control effort.

For completeness, the DeePC formulation used as the benchmark controller \cite{ref9} is briefly summarized below. DeePC constructs predicted input–output trajectories directly from measured data and solves the following optimization problem,
\begin{align}
\min_{u,y,g,\sigma} \sum_{k=1}^{N}
\left(\|y_k\!-\!r_{t+k}\|_Q^2 \!+\! \|u_k\!-\!u^s_{t+k}\|_R^2\right)
\!+\! \lambda_g \|g\|_2^2 \!+\! \lambda_\sigma \|\sigma\|_2^2
\end{align}
subject to
\begin{align}
\begin{bmatrix}
U_p\\
Y_p\\
U_f\\
Y_f
\end{bmatrix} g =
\begin{bmatrix}
u_{\text{ini}}\\
y_{\text{ini}}\\
u\\
y
\end{bmatrix}
+
\begin{bmatrix}
0\\
\sigma\\
0\\
0
\end{bmatrix},
\end{align}
\[
\begin{aligned}
u_k &\in {\mathcal{U}},\quad \forall k\in[1,N], \\
y_k &\in {\mathcal{Y}},\quad \forall k\in[1,N],
\end{aligned}
\]
where $U_p,Y_p,U_f,Y_f$ denote Hankel matrices constructed from the measured input–output data, $u_{\text{ini}}$ and $y_{\text{ini}}$ represent the past trajectories used to initialize the prediction, and $\sigma$ is a slack variable introduced to improve robustness.

Since DKPC and DeePC essentially solve different optimization problems, their cost functions are not directly comparable. To ensure a fair comparison, DKPC and DeePC were evaluated over the same parameter ranges for $q\in \{10,\,100,\,300,\,1000\}$, $r\in \{0.001,\,0.01,\,0.1,\,1\}$, $\lambda_g\in \{1,\,10,\,100,\,1000\}$; the penalty parameter for the slack variable $\sigma$ in DeePC is chosen from $\lambda_{\sigma}\in \{10^{4},\,10^{5},\,10^{6}\}$; All the other settings were kept identical.
Therefore, the DKPC and DeePC controllers are evaluated over 64 and 192 parameter combinations, respectively.

The performance of both controllers is evaluated based on tracking performance $\epsilon$ in \eqref{eq:epsilon} and control effort in $J_u$ \eqref{eq:Ju}. 
Fig.~\ref{fig:tradeoff_frontier} presents the corresponding trade-off in the \(J_u-\epsilon\) plane. The scattered points denote results from all parameter combinations, while the solid curves denote the optimal frontier since the lower-left direction corresponds to both smaller tracking error and lower control effort. The relative position of DKPC's frontier lies closer to the lower-left region than DeePC across the main comparison region, which reflects a better trade-off, i.e., DKPC attains lower tracking error for comparable control effort and, conversely, requires lower control effort for comparable traking error.

\begin{figure}[t]
    \centering
    \includegraphics[width=0.95\linewidth]{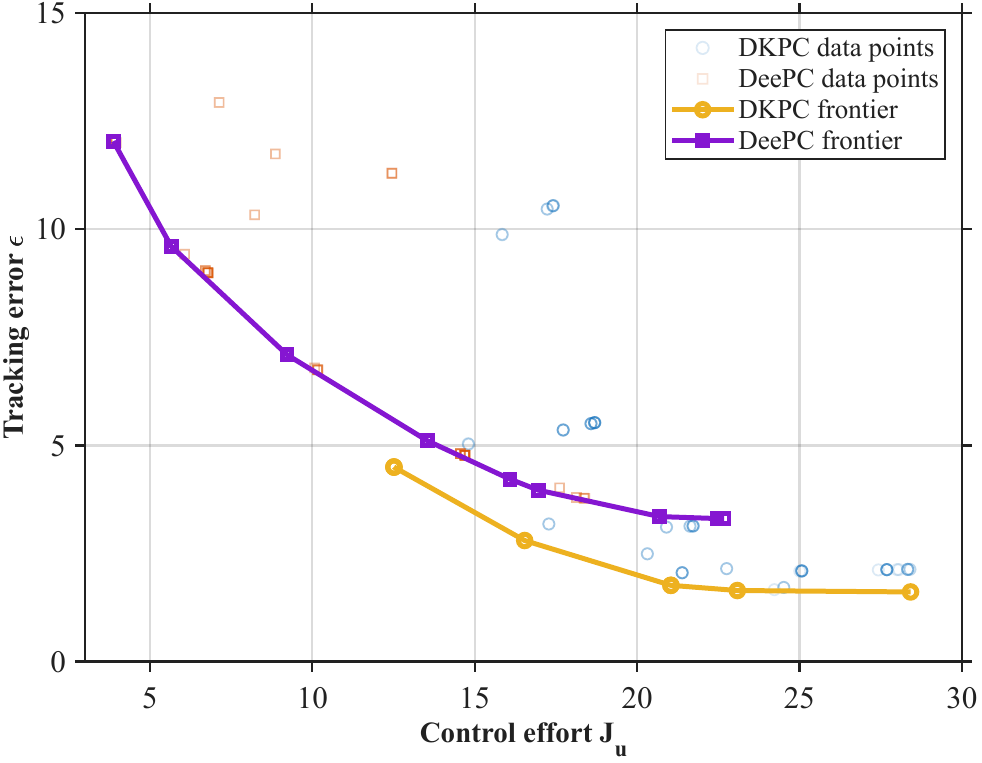}
    \caption{Trade-off between tracking error and control effort for the DKPC and DeePC controllers.}
    \label{fig:tradeoff_frontier}
\end{figure}

Since tracking performance and control effort represent competing objectives, a mixed performance index, $S_{\alpha}$, is introduced to evaluate the trade-off between them,
\begin{equation}
S_{\alpha}
=
\alpha \epsilon
+
(1-\alpha) J_{u},
\label{eq:blended_index}
\end{equation}
where \(\alpha \in [0,1]\) represents the preference between tracking performance and control effort in the evaluation. Larger \(\alpha\) places more emphasis on tracking performance, while smaller \(\alpha\) places more emphasis on control effort.

The parameter \(\alpha\) was uniformly sampled over \([0,1]\). For each \(\alpha\), the index \(S_{\alpha}\) was evaluated for all parameter combinations for each controller, and the case with the minimum index was selected as the best. 

Fig.~\ref{fig:weighted_score_alpha_rev} shows the optimal mixed index as a function of \(\alpha\).
DKPC remains below DeePC over the interval for larger \(\alpha\). 
Note that the control tuning in this study is regulation-oriented, i.e., with tracking performance weighted much more heavily than control effort in the objective functions ($q/r \gg 1$). Fig.~\ref{fig:weighted_score_alpha_rev} shows that DKPC was able to follow such design choice in controller synthesis and achieves a better trade-off corresponding to higher values of \(\alpha\) that emphasize tracking performance.

\begin{figure}[t]
    \centering
    \includegraphics[width=0.95\linewidth]{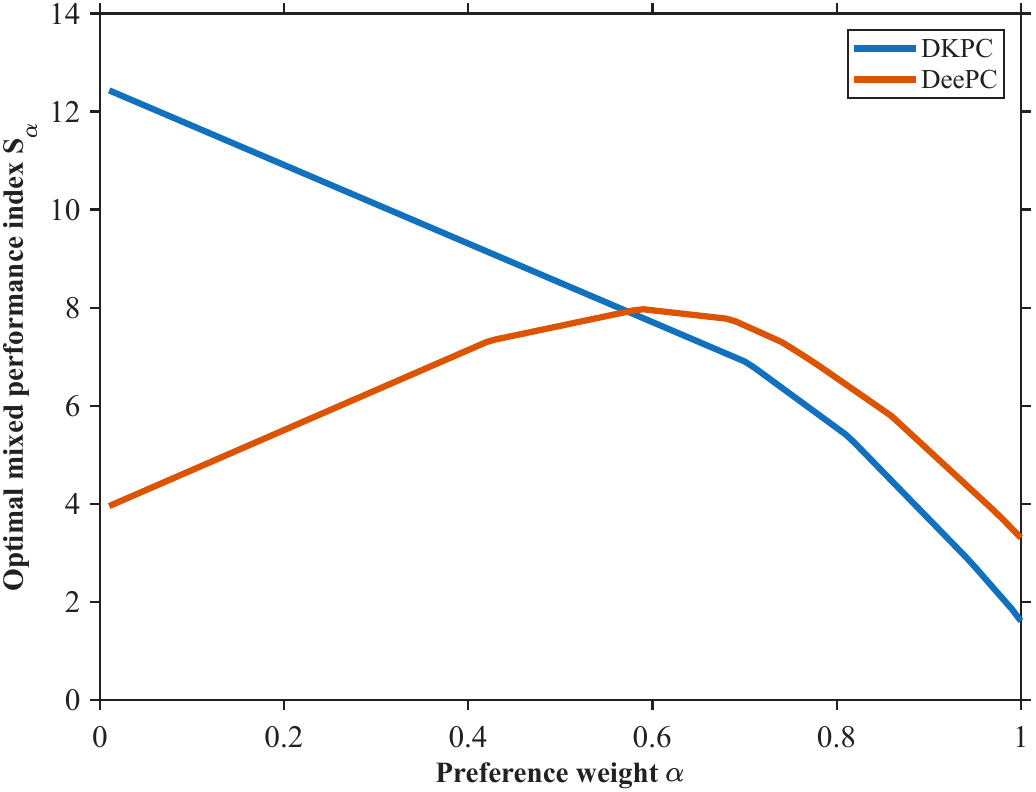}
    \caption{Optimal mixed performance index $S_\alpha$ versus the trade-off parameter $\alpha$ for DKPC and DeePC. The index is computed using min-max normalizations.}
    \label{fig:weighted_score_alpha_rev}
\end{figure}

Overall, these results indicate that DKPC achieves a more favorable trade-off between tracking performance and control effort compared to DeePC over the tested parameter ranges. This improvement is likely due to the Koopman lifting, which provides a richer representation of the system behavior and enables more effective prediction within the data-driven predictive control framework. 
}

\section{Conclusion}
This paper presents a data-driven Koopman predictive frequency control framework to address the challenges brought by black-box models of inverter-based resources (IBRs) in future power systems. 
By combining Koopman theory with Willems’ fundamental lemma, the proposed approach constructs a behavioral model directly from input–output data and formulates a receding-horizon predictive control problem, that balances tracking performance and control effort under explicit input and output constraints.  

Simulation results on the IBR-based IEEE 39-bus system demonstrate that the proposed controller can effectively regulate system frequency and respects bounds on control inputs without requiring an explicit model. The use of lifted Koopman representations and data-consistent trajectory constraints enables efficient convex optimization and provides an interpretable predictive control design. Superior performance was demonstrated as benchmarked against the DeePC control.
These findings highlight the potential of data-driven, model-free predictive control strategies for IBR-dominated power grids. 
Future work will 
extend to larger-scale power systems with heterogeneous IBRs and explore theoretical guarantees on stability. Focus will also be given to the communication and computation needs of the centralized formulation, and the potential extension to decentralized control.

\bibliographystyle{IEEEtran}
\bibliography{ref}

\end{document}